\documentclass[12pt]{article}
\usepackage{latexsym}
\usepackage{amsfonts}
\usepackage{xy}
\input xy
\xyoption{all}

\title{Symmetry Classification of Diatomic Molecular Chains}
\author{S. Lafortune\thanks{Department of Mathematics, University of Arizona, 
P.O. Box 210089, Tucson, Arizona 85721-0089 USA 
(\texttt{lafortus@math.arizona.edu})}
\and S. Tremblay\thanks{Centre de Recherches Math\'ematiques and D\'epartement
de Physique, Universit\'e de Montr\'eal, C.P. 6128, succ. Centre-ville, 
Montr\'eal (QC), H3C 3J7, Canada (\texttt{tremblay@crm.umontreal.ca})} \and 
P. Winternitz\thanks{Centre de Recherches 
Math\'ematiques and D\'epartement de Math\'ematiques et de Statistique, 
Universit\'e de Montr\'eal, 
C.P. 6128, succ. Centre-ville, Montr\'eal (QC), H3C 3J7, Canada 
(\texttt{wintern@crm.umontreal.ca})}}
\date{April 4, 2001}
 
\makeatletter
\@addtoreset{equation}{section}
    
\makeatletter
 
\def\pa {\partial}
\def\ti {\tilde}

\def\eq {\equiv}

\def\al{\alpha}
\def\b{\beta}
\def\ga{\gamma}

\def\ep{\varepsilon}

\def\la{\lambda}

\def\si{\sigma}

\def\be   {\begin{equation}}   \def\ee   {\end{equation}}
\def\ba   {\begin{array}}      \def\ea   {\end{array}}
\def\bea  {\begin{eqnarray}}   \def\eea  {\end{eqnarray}}
\def\bean {\begin{eqnarray*}}  \def\eean {\end{eqnarray*}}

\newcommand{\me}{\mathrm{e}}

\newcommand{\md}{\mathrm{d}}
\newcommand{\mpr}{\mathrm{pr}}
\newcommand{\diag}{\mathrm{diag\, }}

\def\xn{x_n}
\def\xu{x_{n+1}}

\def\yn{y_n}

\def\xin{\xi_n}

\def\etan{\eta_n}

\def\etad{\eta_{n-1}}
\def\lan{\lambda_n}
\def\lau{\lambda_{n+1}}

\def\laon{\lambda_{1,n}}
\def\laou{\lambda_{1,n+1}}
\def\laod{\lambda_{1,n-1}}
\def\latn{\lambda_{2,n}}
\def\latu{\lambda_{2,n+1}}
\def\latd{\lambda_{2,n-1}}
\def\mun{\mu_n}

\def\mud{\mu_{n-1}}
\def\Fn{F_n}

\def\Gn{G_n}

\def\Kn{K_n}

\def\Pn{P_{n}}

\def\pat{\partial_t}
\def\paxn{\partial_{x_n}}
\def\payn{\partial_{y_n}}
\def\omn{\omega_{n}}

\def\zen{\zeta_{n}}

\def\zed{\zeta_{n-1}}

\def\fn{f_n}

\def\gn{g_n}

\def\kn{k_n}

\def\pn{p_n}

\def\sn{\sigma_n}
\def\su{\sigma_{n+1}}
\def\sd{\sigma_{n-1}}
 
\begin{document}
\maketitle

\begin{abstract}
A symmetry classification of possible interactions in a diatomic molecular 
chain is provided. For nonlinear interactions the group of Lie point 
transformations, leaving the lattice invariant and taking solutions into 
solutions, is at most five-dimensional. An example is considered in which 
subgroups of the symmetry group are used to reduce the dynamical 
differential-difference equations to purely difference ones.
\end{abstract}


\newpage

\section{Introduction}

The purpose of this article is to analyze possible interactions in a long 
one-dimensional molecule consisting of two types of atoms. The model we 
consider is a very general one, described by the equations

\be
\ba{rcl}
E_1 &\eq& \ddot{x}_n-\Fn(\xin,t)-\Gn(\etad,t)=0, 
\\*[2ex]
E_2 &\eq& \ddot{y}_n-\Kn(\xin,t)-\Pn(\etan,t)=0,
\label{eq:1.1}
\ea
\ee
where the overdots denote time derivatives and $\xn$, $\yn$ can be 
interpreted as the displacement of the $n$-th atom 
of type $X$ or $Y$, respectively, from their equilibrium positions. We define
\be
\xin\eq \yn-\xn,\ \ \ \ \ \ \etan\eq \xu-\yn
\label{eq:1.2}
\ee
and $t$ is time. The functions $\Fn,\, \Gn,\, \Kn$ and $\Pn$ are as yet 
unspecified smooth functions. Indeed, our aim is to classify such systems 
according to the Lie point symmetries that they allow, that is, to classify 
these functions 
$\Fn,\, \Gn,\, \Kn$ and $\Pn$.

The assumptions built into the model are:

\begin{enumerate}

\item The atoms of type $X$ and $Y$ alternate along a fixed uniform 
one-dimensional chain with positions labeled by the integers $n$ \\ 
(see Figure 1).

\item Only nearest neighbor interactions are considered, i.e. the atom $X_n$ 
interacts only with $Y_{n-1}$ and $Y_n$ and $Y_n$ interacts only with $X_n$
 and $X_{n+1}$ (see Figure~1).

\item The system is invariant under a uniform translation of all atoms in the
 molecule and also under a Galilei transformations of the chain.

\item The systems is strongly coupled, i.e. we assume
\be
\frac{\pa\Fn}{\pa\xin}\ne 0,\ \ \frac{\pa \Gn}{\pa\etad}\ne 0,
\ \ \frac{\pa\Kn}{\pa\xin}\ne 0,\ \ \frac{\pa \Pn}{\pa\etan}\ne 0.
\label{eq:1.3}
\ee

\item In the bulk of the article we assume that the interactions are 
nonlinear, i.e. at least one of the four functions $\Fn,\, \Gn,\, \Kn$ or 
$\Pn$ 
depends nonlinearly on the argument $\xi$ or $\eta$, respectively. The linear 
case will be treated separately.

\item A discrete symmetry is built into the model. Indeed, the two equations 
(\ref{eq:1.1}) are permuted by the transformation

\be
\ba{c}
\xn \longrightarrow \yn,\ \ \ \ \yn \longrightarrow \xu, 
\\*[2ex]
\Fn \longrightarrow \Pn,\ \ \ \ \Gn \longrightarrow \Kn,\ \ \ \ K_{n-1} 
\longrightarrow \Gn,\ \ \ \ P_{n-1} \longrightarrow \Fn.
\label{eq:1.4}
\ea
\ee
   
\end{enumerate}

Models of this type have many applications in classical mechanics, in
 molecular physics, or mathematical biology \cite{1,2,3}. In applications, 
the form of the
 functions in eq.(\ref{eq:1.1}) are usually {\em a priori} fixed.

The formalism used in this article is the one called ``intrinsic method'' in 
earlier articles \cite{4,5}. It has already been applied to monoatomic 
molecular chains \cite{6} and to a model with two species, or two types of 
atoms, distributed along a double chain \cite{7}.

In this approach the dependent variables $x$ and $y$ depend on one discrete 
variable $n$ and one continuous variable $t$. Symmetry transformations, 
taking solutions into solutions, act on the variables $x,y$ and $t$, not 
however on the lattice variable $n$. The Lie algebra of the symmetry group is 
realized by vector fields of the form

\be
\hat X=\tau(\xn,\yn,t)\pa_t+\phi_n(\xn,\yn,t)\pa_{\xn}+
\psi_n(\xn,\yn,t)\pa_{\yn}.
\label{eq:1.5}
\ee
The functions $\tau, \phi_n$ and $\psi_n$ are determined from the requirement 
that the second prolongation of the vector field $\hat X$ should annihilate 
equations (\ref{eq:1.1}) on their solution surface. Explicitly we have 
\cite{4,5,6,7}

\be
\ba{rcl}
\mpr^{(2)}\hat{X}&=& \tau(t,\xn,\yn)\pa_t+
\sum\limits_{k=n-1}^{n+1}\phi_k(t,\xn,\yn)\pa_{x_k}
\\*[2ex]
&+&\sum\limits_{k=n-1}^{n+1}\psi_k(t,\xn,\yn)\pa_{y_k}+
\phi_{n}^{tt}\pa_{\ddot{x}_n}+
\psi_{n}^{tt}\pa_{\ddot{y}_n}
\label{eq:1.6}
\ea
\ee
with
\be
\ba{rcl}
\phi_{n}^{tt} &=& D_{t}^{2}\phi_n-(D_{t}^{2}\tau)\, \dot{x}_n
-2(D_{t}\tau)\, \ddot{x}_n,
\\*[2ex]
\psi_{n}^{tt} &=& D_{t}^{2}\psi_n-(D_{t}^{2}\tau)\, \dot{y}_n
-2(D_{t}\tau)\, \ddot{y}_n
\label{eq:1.7}
\ea
\ee
($D_t$ is the total time derivative). In eq.(\ref{eq:1.6}) 
we have spelled 
out only those terms which act on eq.(\ref{eq:1.1}).

The use of this formalism is not obligatory. Indeed, the group transformations
 can also act on the lattice \cite{8,9,10,11} and generalized symmetries
 can be very useful \cite{12}. In this article we restrict ourselves to the 
intrinsic formalism, described above.

The present article is organized as follows. In Section~2 we establish the 
general form of the vector fields (\ref{eq:1.5}) that realize the symmetry
 algebra of eq.(\ref{eq:1.1}). We also derive the determining equations for 
the symmetries and introduce a ``group of allowed transformations''. Allowed
 transformations take equations of the type (\ref{eq:1.1}) into other 
equations of the same 
type. They can change the functions $\Fn,\, \Gn,\, \Kn$ and $\Pn$ into other 
functions of the same arguments. As in previous articles, we classify equations
 into symmetry classes under the action of allowed transformations 
\cite{6,7,13,14}. We also establish that equations (\ref{eq:1.1}) are 
invariant under a two-dimensional Abelian group for any functions 
$\Fn,\ldots,\Pn$. Section~3 is devoted to Abelian symmetry algebras. We denote
 them $A_{j,k}$ where $A$ means Abelian, $j$ denotes the dimension and 
$k=1,2,3,\ldots$ enumerates algebras of the same dimension. For each 
interaction we list only the maximal symmetry algebra. Section~4 is devoted 
to nilpotent symmetry algebras, denoted by $N_{j,k}$ with the same conventions 
as in Section~3. In Section~5 we find all solvable symmetry algebras with 
non-Abelian nilradicals ($SN_{j,k}$). In Section~6 those with Abelian 
nilradicals ($SA_{j,k}$). All nonsolvable symmetry algebras are listed in 
Section~7 ($NS_{j,k}$). In Sections 3 to 7 we consider only nonlinear 
interactions. Symmetries of the linear case are discussed in Section~8. 
Conclusions and some applications of the symmetries are summed up in the 
final Section~9.

\section{Determining equations and allowed \\ transformations}

The algorithm for finding the symmetry algebra of eq.(\ref{eq:1.1}) is

\be
\mpr\hat{X}\, E_{a}\left|_{E_b=0}=0,\right.\ \ \ \ \ \ a=1,2,\ \ b=1,2.
\label{eq:2.1}
\ee
The coefficients of all terms of the type $\dot{x}_{n}^{p}\, \dot{y}_{n}^{q}$
 must vanish independently and we find that the vector field (\ref{eq:1.5})
 must actually have the form

\be
\hat X=\tau(t)\pat + 
\left[\left(a+\frac{\dot{\tau}(t)}{2}\right)\xn+\lan(t)\right]\paxn + 
\left[\left(a+\frac{\dot{\tau}(t)}{2}\right)\yn+\mun(t)\right]\payn,
\label{eq:2.2}
\ee
where $a$ is a constant and $\lan(t), \mun(t)$ and $\tau(t)$ are functions of 
the indicated variables. This form (\ref{eq:2.2}) is valid for any 
interactions $\Fn,\, \Gn,\, \Kn$ and $\Pn$ in eq.(\ref{eq:1.1}). Moreover, we
 have

\be
\tau=\tau_0+\tau_1\, t+\tau_2\, t^2,
\label{eq:2.3}
\ee
where $\tau_0, \tau_1$ and $\tau_2$ are constants.

The constants $a$, $\tau_i$ and the functions $\lan(t)$ and $\mun(t)$ are 
subject to two further determining equations that involve the interaction 
functions explicitly. They are

\be
\ba{l}
\ddot{\la}_n+\left(a-\frac{3}{2}\dot{\tau}\right)(\Fn+\Gn)+
\left[\lan-\mun-\left
(a+\frac{\dot{\tau}}{2}\right)\xin\right]F_{n,\xin}
\\*[2ex]
+\left[\mud-\lan-
\left(a+\frac{\dot{\tau}}{2}\right)\etad\right]G_{n,\etad}-\tau
(F_{n,t}+G_{n,t})=0,
\ea
\label{eq:2.4}
\ee
\vspace{.5cm}
\be
\ba{l}
\ddot{\mu}_n+\left(a-\frac{3}{2}\dot{\tau}\right)(\Kn+\Pn)+
\left[\lan-\mun-\left
(a+\frac{\dot{\tau}}{2}\right)\xin\right]K_{n,\xin}
\\*[2ex]
+\left[\mun-\lau-\left(a+\frac{\dot{\tau}}{2}\right)\etan\right]P_{n,\etan}-
\tau(K_{n,t}+P_{n,t})=0.
\ea
\label{eq:2.5}
\ee

Our task is to perform a complete analysis of eq.(\ref{eq:2.4}) and 
(\ref{eq:2.5}). Conceptually, this is very similar to the problem considered 
in Ref.~7. However, the functions figuring in eq.(\ref{eq:1.1}) are less
 general than those of Ref.~7, hence the computations are simpler.

We shall classify the equations of type (\ref{eq:1.1}) into equivalence 
classes under the action of a group of ``allowed transformations''. These are 
transformations of the form

\be
\xn=\Phi_n(\ti{x}_n,\ti{y}_n,\ti{t}),\ \ \ \ \ \ \ \ 
\yn=\Psi_n(\ti{x}_n,\ti{y}_n,\ti{t}),\ \ \ \ \ \ \ \ t=T(\ti{t})
\label{eq:2.6}
\ee
that transform equations (\ref{eq:1.1}) into equations of the same form, but 
do not preserve the functions on the right hand side of eq.(\ref{eq:1.1}). The 
requirement that no first derivatives should appear and that the transformed 
functions $\ti{F}_n$ and  $\ti{K}_n$ should depend only on $\ti{\xi}_n$ and 
$\ti{t}$, $\ti{G}_n$ and  $\ti{P}_n$ only on $\ti{t}$ and $\ti{\eta}_{n-1}$ or 
$\ti{\eta}_{n}$, respectively, implies that the transformations actually have
 the form

\bea
\left(
\ba{c}
\xn(t) \\
\yn(t)
\ea
\right) = q\, \dot{\ti t}^{-1/2}\left(
\ba{c}
\tilde{x}_{n}(\ti{t}) \\
\tilde{y}_{n}(\ti{t})
\ea
\right) +
\left(
\ba{c}
\al_n(t) \\
\b_{n}(t)
\ea
\right),
\label {eq:2.7}
\\*[2ex]
\ti{t}=\frac{c_{1}\, t+c_{2}}{c_{3}\, t+c_{4}},\ \ \ \ \ \ 
c_{1}c_{4}-c_{2}c_{3}=1,\ \ \ \ \ \ q\ne 0,
\label{eq:2.8}
\end{eqnarray}
where $q, c_1, \ldots, c_4$ are constants and $\al_{n}$ and $\b_{n}$ are 
arbitrary functions of $n$ and $t$.

The transformed system is

\be
\ba{rcl}
\ddot{\ti{x}}_n(\ti{t})&=& \ti{F}_n(\ti{\xi}_n,\ti{t})+
 \ti{G}_n(\ti{\eta}_{n-1},\ti{t}),
\\*[2ex]
\ddot{\ti{y}}_n(\ti{t})&=& \ti{K}_n(\ti{\xi}_n,\ti{t})+
 \ti{P}_n(\ti{\eta}_{n},\ti{t}),
\label{eq:2.9}
\ea
\ee
with

\be
\left(
\ba{c}
\ti{F}_{n}+\ti{G}_{n} \\
\ti{K}_{n}+\ti{P}_{n}
\ea
\right) = \frac{\dot{\ti t}^{-3/2}}{q}
\left[
\left(
\ba{c}
F_n(\xin,t)+G_n(\etad,t) \\
K_n(\xin,t)+P_n(\etan,t)
\ea
\right) - 
\left(
\ba{c}
\ddot{\al}_{n}(t) \\
\ddot{\b}_{n}(t)
\ea
\right)
\right],
\label{eq:2.10}
\end{equation}
where
\bea
\xin&=&\yn-\xn=q\, \dot{\ti t}^{-1/2}(\ti{x}_{n}-\ti{y}_{n})+\al(t)-\b(t),
\label{eq:2.11}
\\*[2ex]
\etan&=&\xu-\yn=q\, \dot{\ti 
t}^{-1/2}(\ti{x}_{n+1}-\ti{y}_{n})+\al_{n+1}(t)-\b_{n}(t), 
\label{eq:2.12}
\\*[2ex]
t&=&\frac{c_{4}\, \ti{t}-c_{2}}{-c_{3}\, \ti{t}+c_{1}}.
\label{eq:2.13}
\end{eqnarray}

The vector field $\hat X$ of eq.(\ref{eq:2.2}) is transformed into a similar 
field with

\bea
\ti{\tau}(\ti{t})&=&\tau(t(\ti{t}))\, \dot{\ti{t}},\ \ \ \ \ \ \ti{a}=a, 
\label{eq:2.15}
\\*[2ex]
\left(
\ba{c}
\ti{\la}_{n}(\ti{t}) \\
\ti{\mu}_{n}(\ti{t})
\ea
\right) &=& \frac{\dot{\ti{t}}^{1/2}}{q}\left[(a+\frac{\dot{\tau}}{2})
\left(
\ba{c}
\al_{n} \\
\b_{n}
\ea
\right) - \tau
\left(
\ba{c}
\dot{\al}_{n} \\
\dot{\b}_{n}
\ea
\right) +
\left(
\ba{c}
\la_{n} \\
\mu_{n}
\ea
\right)\right].
\label{eq:2.16}
\end{eqnarray}

The transformed functions and constants must satisfy the same determining 
equations (\ref{eq:2.4}) and (\ref{eq:2.5}).

As mentioned in the Introduction, translational and Galilei invariance are 
built into the model. That is easy to check. Indeed $\lan=\mun=1$, $a=0$, 
$\tau(t)=0$ and $\lan=\mun=t$, $a=0$, $\tau(t)=0$ are solutions of 
eq.(\ref{eq:2.4}) and (\ref{eq:2.5}) for $\Fn,\, \Gn,\, \Kn$ and $\Pn$ 
arbitrary.
 No other symmetries exist, unless some constraints on the interactions are 
imposed.

We shall use the allowed transformations to simplify the vector fields that
 occur. In particular the coefficient $\tau(t)$ of a given vector field can 
be transformed into one of the following expressions: $0$, $1$, $t$ or $t^2+1$.

Our strategy will be to first find all Abelian symmetry algebras, then all 
nilpotent (non-Abelian) ones. Once these are known, we can determine all
 solvable ones, having the corresponding Abelian, or nilpotent ones as
 nilradicals \cite{15}. Finally, all nonsolvable symmetry algebras will be 
determined, making use of their Levi decomposition \cite{15}.

Any symmetry algebra will contain the algebra

\be
\ba{lr}
A_{2,1}: & \hat{X}_1=\paxn+\payn,\ \ \ \ \hat{X}_2=t(\paxn+\payn),
\ea
\label{eq:2.17}
\ee
as a subalgebra. Allowed transformations leave the algebra (\ref{eq:2.17}) 
invariant. Any further element of the symmetry algebra can be transformed 
into one of the following ones

\bea
\hat{Y}_1 &=& \pa_t+a(\xn\, \pa_{\xn}+\yn\, \pa_{\yn}),\ \ \ \ \ \ \ \ a=0,1,
\label{eq:2.18} \\*[2ex]
\hat{Y}_2 &=& t\, \pa_t+
(a+\frac{1}{2})(\xn\, \pa_{\xn}+\yn\, \pa_{\yn}),
\label{eq:2.19} \\*[2ex]
\hat{Y}_3 &=& (t^2+1)\, \pa_t+(a+t)(\xn\, \pa_{\xn}+\yn\, \pa_{\yn}),
\label{eq:2.20} \\*[2ex]
\hat{Y}_4 &=& \lan(t)(\pa_{\xn}+\pa_{\yn}),\ \ \ \ \ \ \ \ 
\ddot{\la}_n\ne 0,\ \ \lau\ne \lan,
\label{eq:2.21} \\*[2ex]
\hat{Y}_5 &=& \lan(t)\, \pa_{\xn}+\lau(t)\, \pa_{\yn},\ \ \ \ \ \ \ \ 
\ddot{\la}_n\ne 0,\ \ \lau\ne \lan.
\label{eq:2.22}
\eea
The interactions that allow these additional terms can easily be determined 
from equations (\ref{eq:2.4}) and (\ref{eq:2.5}). Once this is done, we 
determine whether the considered interactions allows further symmetries. For 
each interaction, we shall only list the maximal symmetry algebra allowed, 
not lower-dimensional subalgebras.

\section{Abelian symmetry algebras}

The lowest dimensional maximal symmetry algebra is $A_{2,1}$ of 
eq.(\ref{eq:2.17}), present for any interactions in eq.(\ref{eq:1.1}). This
 algebra can be enlarged into a higher dimensional Abelian algebra by adding
 elements of the type (\ref{eq:2.21}) or (\ref{eq:2.22}). The determining 
equations for a nonlinear system allow at most four commuting symmetry 
generators. Moreover, the three-dimensional symmetry algebras are never
 maximal.

Finally, we obtain two different four-dimensional Abelian symmetry algebras 
together with the interactions that allow them. They are

\[
\ba{ll}
A_{4,1} & \hat{X}_1=\laon(t)(\paxn+\payn),\ \ \ \ 
\hat{X}_2=\latn(t)(\paxn+\payn), \\*[2ex]
& \hat{X}_3=\paxn+\payn,\ \ \ \ \hat{X}_4=t(\paxn+\payn), \\*[2ex]
& \Fn=\Fn(\xin,t),\ \ \ \ \ 
\Gn=\frac{\displaystyle\ddot{\la}_{1,n}}{\displaystyle\laon-\laod}\etad, 
\\*[2ex]
& \Kn=\Kn(\xin,t),\ \ \ \ \ 
\Pn=\frac{\displaystyle\ddot{\la}_{1,n}}{\displaystyle\laou-\laon}\etan,
\\*[2ex]
& \ddot{\la}_{1,n}\ne 0,\ \ \ \ \ddot{\la}_{2,n}\ne 0,\ \ \ \ \laou\ne \laon, 
\\*[2ex]
&  \frac{\displaystyle\ddot{\la}_{2,n}}{\displaystyle\ddot{\la}_{1,n}}=
\frac{\displaystyle\latn-\latd}{\displaystyle\laon-\laod}=
\frac{\displaystyle\latu-\latn}{\displaystyle\laou-\laon}. \\
\\
A_{4,2} & \hat{X}_1=\laon(t)\paxn+\laou(t)\payn,\ \ \ \ 
\hat{X}_2=\latn(t)\paxn+\latu(t)\payn, \\*[2ex]
& \hat{X}_3=\paxn+\payn,\ \ \ \ \hat{X}_4=t(\paxn+\payn),  \\*[2ex]
&\Fn=\frac{\displaystyle\ddot{\la}_{1,n}}{\displaystyle\laou-\laon}\xin,
\ \ \ \ \Gn=\Gn(\etad,t), \\*[2ex]
&\Kn=\frac{\displaystyle\ddot{\la}_{1,n+1}}{\displaystyle\laou-\laon}\xin,
\ \ \ \ \Pn=\Pn(\etan,t), \\*[2ex] 
& \ddot{\la}_{1,n}\ne 0,\ \ \ \ \ddot{\la}_{2,n}\ne 0,\ \ \ \ \laou\ne \laon, 
\\*[2ex]
& \frac{\displaystyle\ddot{\la}_{2,n}}{\displaystyle\ddot{\la}_{1,n}}=
\frac{\displaystyle\latn-\latd}{\displaystyle\laon-\laod}=
\frac{\displaystyle\latu-\latn}{\displaystyle\laou-\laon}. \\
\label{eq:4.2}
\ea
\]
The algebras $A_{4,1}$ and $A_{4,2}$ are actually related by the discrete 
symmetry (\ref{eq:1.4}). Algebra $A_{4,1}$ is transformed into $A_{4,2}$ by 
the substitutions

\be
\ba{c}
\Fn(\xin) \longrightarrow \Pn(\etan),\ \ \ \ 
\Gn(\etad) \longrightarrow \Kn(\xin),
\\*[2ex]
K_{n-1}(\xi_{n-1}) \longrightarrow \Gn(\etad),\ \ \ \  
P_{n-1}(\eta_{n-1}) \longrightarrow \Fn(\xin),
\\*[2ex]
\sn(t)\, \pa_{\xn} \longrightarrow \su(t)\, \pa_{\yn},\ \ \ \  
\sn(t)\, \pa_{\yn} \longrightarrow \sn(t)\, \pa_{\xn}.
\label{eq:4.3}
\ea
\ee
The functions $\laon(t)$ and $\latn(t)$ in algebras $A_{4,1}$, $A_{4,2}$ 
satisfy the equations

\be
\frac{\ddot{\la}_{2,n}}{\ddot{\la}_{1,n}}=\frac{\latn-\latd}{\laon-\laod}=
\frac{\latu-\latn}{\laou-\laon}.
\label{eq:4.4}
\ee
These equations can be solved and we obtain
\be
\ba{c}
\laon=f(t)\latn+g(t),\ \ \ \ 
\latn=\frac{\displaystyle\ga_n}{\displaystyle\dot{f}(t)^{1/2}}-
\frac{\displaystyle1}{\displaystyle2\, 
\dot{f}(t)^{1/2}}\displaystyle\int_{t_0}^{t}
\frac{\displaystyle\ddot{g}(s)}{\displaystyle\dot{f}(s)^{1/2}}\md s,
\\*[2ex]
\dot{f}(t)\ne 0,\ \ \ \ \ \ \ \ \ga_{n+1}\ne \ga_n,
\label{eq:4.5}
\ea
\ee
where $f(t)$, $g(t)$ are arbitrary smooth functions of $t$ and $\ga_n$ is an 
arbitrary function of $n$.

Notice that the quantities $\laon(t)$ and $\latn(t)$ (or $f(t)$, $g(t)$ and 
$\ga_n$) figure explicitly in the interaction functions $\Gn$ and $\Pn$ of 
$A_{4,1}$, or respectively in $\Fn$ and $\Kn$ of $A_{4,2}$. The two algebras
 are thus indeed four-dimensional and completely specified.

\section{Nilpotent non-Abelian symmetry algebras}

Nilpotent Lie algebras exist for all dimensions $\dim\, L\geq3$. For 
$\dim\, L=3$ only 
one type exists, namely the Heisenberg algebra. It has a two-dimensional 
Abelian ideal. Maximality requires that this ideal be the algebra $A_{2,1}$ of
 eq.(\ref{eq:2.17}). The Heisenberg algebra is obtained by adding the 
operator $\hat{T}=\pa_t$. We then calculate the interaction allowing this 
symmetry algebra, and obtain

\[
\ba{ll}
N_{3,1} & \hat{X}_1=\paxn+\payn,\ \ \ \ \hat{X}_2=t(\paxn+\payn),\ \ \ \ 
\hat{T}=\pat, \\*[2ex]
& \Fn=\fn(\xin),\ \ \ \ \Gn=\gn(\etad), \\*[2ex]
& \Kn=\kn(\xin),\ \ \ \ \Pn=\pn(\etan). 
\ea
\]
We mention that this algebra is invariant under the substitution 
(\ref{eq:4.3}).

Every nilpotent non-Abelian Lie algebra contains the Heisenberg algebra as a 
subalgebra. We can hence proceed by adding further operators to $N_{3,1}$. 
Moreover, they can only be added to the Abelian ideal. The determining 
equations (\ref{eq:2.4}), (\ref{eq:2.5}) allow us to add at most two 
operators. Maximality requires that we add precisely two. We thus obtain two 
mutually isomorphic five-dimensional nilpotent Lie algebras with 
four-dimensional Abelian ideals, namely 

\[
\ba{ll}
N_{5,1} & \hat{X}_1=\paxn+\payn,\ \ \ \ \hat{X}_2=t(\paxn+\payn),\ \ \ \ 
\hat{T}=\pat, \\*[2ex]
& \hat{X}_3=(\sn+t^2)(\paxn+\payn),\ \ \ \ \hat{X}_4=(\sn 
t+\frac{t^3}{3})(\paxn+\payn), \\*[2ex]
&\Fn=\fn(\xin),\ \ \ \ \Gn=\frac{\displaystyle2}{\displaystyle\sn-\sd}\etad, 
\\*[2ex]
&\Kn=\kn(\xin),\ \ \ \ \Pn=\frac{\displaystyle2}{\displaystyle\su-\sn}\etan,
\ \ \ \ \su\ne \sn. 
\ea
\]

The second algebra $N_{5,2}$ is obtained from $N_{5,1}$ by the substitution
 (\ref{eq:4.3}). We mention that the interactions allowing the symmetry 
algebra $N_{5,1}$ are special cases of those allowing the Abelian algebra 
$A_{4,1}$. Similarly for $N_{5,2}$ and $A_{4,2}$.

\section{Solvable nonnilpotent symmetry algebras with non-Abelian nilradicals}

A solvable Lie algebra $L$ always has a uniquely defined maximal nilpotent 
ideal, the nilradical $NR(L)$ \cite{15}. If a solvable symmetry algebra of 
the system (\ref{eq:1.1}) has a non-Abelian nilradical, it must be $N_{3,1}$, 
$N_{5,1}$ or $N_{5,2}$ of Section~4, or a four-dimensional subalgebra of  
$N_{5,1}$ or $N_{5,2}$.

The determining equations (\ref{eq:2.4}) and (\ref{eq:2.5}) do not allow any 
extension of the four and five-dimensional nilpotent symmetry algebras to 
solvable ones.

The Heisenberg algebra $N_{3,1}$, on the other hand, leads to three different
 four-dimensional solvable symmetry algebras. The Lie algebras are given by 
four basis elements, $\hat{X}_1, \hat{X}_2$ and $\hat{T}$ of $N_{3,1}$ and an 
additional operator $\hat{Y}$. Below we list these elements $\hat{Y}$ 
together with the invariant interactions that allow the corresponding
 symmetry groups. In each case we present a matrix $A$ defining the action of 
 $\hat{Y}$ on the nilradical $N_{3,1}$.

\[
\ba{ll}
SN_{4,1} & \hat{Y}=t\pat+(a+\frac{1}{2})(\xn \paxn+\yn \payn), \\*[2ex]
& \Fn=(\xin)^{\frac{2a-3}{2a+1}}\fn,\ \ \ \ 
\Gn=(\etad)^{\frac{2a-3}{2a+1}}\gn, 
\\*[2ex]
& \Kn=(\xin)^{\frac{2a-3}{2a+1}}\kn,\ \ \ \ 
\Pn=(\etan)^{\frac{2a-3}{2a+1}}\pn, 
\\*[2ex]
& A=\diag(a+\frac{1}{2}\, ,\, 1\, ,\, a-\frac{1}{2}),
\ \ \ \ a\ne -\frac{1}{2}\, ,\, \frac{3}{2}. \\
\\
SN_{4,2} & \hat{Y}=t\pat+(2\xn+t^2)\paxn+(2\yn+t^2)\payn, \\*[2ex]
& \Fn=\fn+\frac{1}{2}\ln(\xin),\ \ \ \ \Gn=\frac{1}{2}\ln(\etad), \\*[2ex] 
& \Kn=\kn+\frac{1}{2}\ln(\xin),\ \ \ \ \Pn=\frac{1}{2}\ln(\etan), \\*[2ex] 
& A=
\left(
\ba{ccc}
2 & 0 & 0     \\
0     & 1 & 2     \\
0     & 0  & 1 \\
\ea
\right). \\
\\
SN_{4,3} & \hat{Y}=t\pat+\si_{1n}\paxn+\si_{2n}\payn, \\*[2ex]
& \Fn=\fn\exp\left(\frac{\displaystyle2\xin}{\displaystyle\si_{1,n}-\si_{2,n}}
\right),
\ \ \ \  
\Gn=\gn\exp\left(\frac{\displaystyle-2\etad}
{\displaystyle\si_{1,n}-\si_{2,n-1}}\right), \\*[2ex]
& \Kn=\kn\exp\left(\frac{\displaystyle2\xin}{\displaystyle\si_{1,n}-\si_{2,n}}
\right),
\ \ \ \  
\Pn=\pn\exp\left(\frac{\displaystyle-2\etan}
{\displaystyle\si_{1,n+1}-\si_{2,n}}\right), \\*[2ex]
& A=\diag(0\, ,\, 1\, ,\, -1),\ \ \ \ \si_{1,n}\ne\si_{2,n},\ \ \ \ 
\si_{1,n+1}\ne\si_{2,n}.
\ea
\]

The quantities $\fn, \gn, \pn, \kn, \si_{1,n}$ and $\si_{2,n}$ depend on $n$
 alone.

The transformation (\ref{eq:4.3}) does not lead to any new algebras or 
interactions. In the case of the algebra $SN_{4,3}$ we may have 
$\si_{2,n+1}=\si_{2,n}$. Then $\si_2$ can be transformed into 
$\si_{2,n}=\si=0$. Similarly, for $\si_{2,n+1}\ne \si_{2,n}$, but 
$\si_{1,n+1}=\si_{1,n}\eq \si$ we can transform into $\si=0$.

\section{Solvable nonnilpotent symmetry algebras with Abelian nilradicals}

A large number of symmetry algebras of the system (\ref{eq:1.1}) is of this 
type. To identify and classify them, we use several known results on the 
structure of solvable Lie algebras \cite{15}.

\begin{enumerate}

\item The nilradical $NR(L)$ is unique and its dimension satisfies
\be
\dim\ NR(L)\geq \frac{1}{2}\ \dim\ L.
\end{equation}

\item Any solvable Lie algebra $L$ can be written as the algebraic sum of the
 nilradical $NR(L)$ and a complementary linear space $F$, i.e. 
$L=F \dot{+} NR(L)$.

\item The derived algebra is contained in its nilradical: $[L,L]\subseteq 
NR(L)$.

\item For an Abelian nilradical $\{\hat{X}_1,\ldots,\hat{X}_n\}$, the 
commutation relations can be written as
\be
[\hat{X}_i,\hat{Y}_k]=(A_{k})_{ij}\hat{X}_{j},\ \ \ \ [A_k,A_\ell]=0,\ \ \ \ 
[\hat{Y}_i,\hat{Y}_k]=c_{ik}^{\ell} \hat{X}_\ell,\ \ \ \ 
[\hat{X}_i,\hat{X}_k]=0,
\label{eq:rcsa}
\end{equation}
where the elements $\hat{Y}_k$ are the nonnilpotent elements (outside the 
nilradical). The matrices $A_{k}$ commute and are linearly nilindependent 
(i.e. no nontrivial linear combination of them is a nilpotent matrix). If 
only one element $\hat Y$ outside the nilradical exists, the nonnilpotent 
matrix $A$ can be taken in Jordan canonical form.

\end{enumerate}
In our case we can add that the Abelian nilradical must be one of the 
algebras found in Section~3. In principle, the nilradical could be a 
three-dimensional subalgebra of $A_{4,1}$, or $A_{4,2}$, containing $A_{2,1}$
 as a subalgebra. However,
it turns out that all choices of this type lead to symmetry algebras that
are not maximal for the interactions that they allow.

The following solvable symmetry algebras occur.

\subsection{$\dim\, NR(L)=2$}

The only two-dimensional nilradical that leads to solvable Lie 
algebras that are maximal for the obtained interaction is $A_{2,1}$. 
The solvable Lie 
algebras are always three dimensional. A basis for them consists of 
$\hat{X}_1$ and 
$\hat{X}_2$ of eq.(\ref{eq:2.17}) and an additional element $\hat{Y}$, given
 below. In each case we give the matrix $A$ representing the action of 
$\hat{Y}$ on the nilradical.

\[
\ba{ll}
SA_{3,1} & \hat{X}_1=\paxn+\payn,\ \ \ \ 
\hat{X}_2=t(\paxn+\payn),\ \ \ \ \hat{Y}=\pat+\xn \paxn+\yn \payn, \\*[2ex]
& \Fn=\xin \fn(\omn),\ \ \ \ \Gn=\etad \gn(\zed), \\*[2ex]
& \Kn=\xin \kn(\omn),\ \ \ \ \Pn=\etan \pn(\zen), \\*[2ex]
& \omn= \xin \me^{-t},\ \ \ \zen=\etan \me^{-t},\ \ \ \ A=\left(
\ba{cc}
1 & -1 \\
0 & 1 \\
\ea
\right). \\
\\
SA_{3,2} & \hat{X}_1=\paxn+\payn,\ \ \ \ 
\hat{X}_2=t(\paxn+\payn),\ \ \ \ \hat{Y}=t\pat+
(a+\frac{1}{2})(\xn \paxn+\yn \payn), \\*[2ex]
& \Fn=t^{-2}\xin \fn(\omn),\ \ \ \ \Gn=t^{-2}\etad \gn(\zed), \\*[2ex]
& \Kn=t^{-2}\xin \kn(\omn),\ \ \ \ \Pn=t^{-2}\etan \pn(\zen), \\*[2ex]
& \omn= \xin t^{-(a+\frac{1}{2})},\ \ \ 
\zen=\etan t^{-(a+\frac{1}{2})},\ \ \ \ 
A=\diag(a-\frac{1}{2}\, ,\, a+\frac{1}{2}).\\
\ea
\]
\[
\ba{ll}
SA_{3,3} & \hat{X}_1=\paxn+\payn,\ \ \ \ 
\hat{X}_2=t(\paxn+\payn),\ \ \ \ 
\hat{Y}=(t^2+1)\pat+(a+t)(\xn \paxn+\yn \payn), \\*[2ex]
& \Fn=(t^2+1)^{-2}\xin \fn(\omn),\ \ \ \ \Gn=(t^2+1)^{-2}\etad \gn(\zed), 
\\*[2ex]
& \Kn=(t^2+1)^{-2}\xin \kn(\omn),\ \ \ \ \Pn=(t^2+1)^{-2}\etan \pn(\zen), 
\\*[2ex]
& \omn= \xin (t^2+1)^{-1/2}\exp[-a\arctan(t)],\ \ \ 
\zen=\etan (t^2+1)^{-1/2}\exp[-a\arctan(t)] \\*[2ex]
& A=\left(
\ba{cc}
a & -1 \\
1 & a \\
\ea
\right). \\
\ea
\]
These three algebras are nonisomorphic (since the corresponding matrices $A$
 are not mutually conjugate). Each of these three cases is self conjugate
 under the substitution (\ref{eq:4.3}).

\subsection{$\dim\, NR(L)=4$}

The nilradical could be three-dimensional, however the obtained solvable Lie 
algebra is never maximal. We only need to deal with four-dimensional Abelian 
ideals of the form $A_{4,1}$ and $A_{4,2}$. An extension to a solvable Lie 
algebra is only possible for special cases of the functions $\laon(t)$ and 
$\latn(t)$ figuring in the vector fields and interactions. Below we list all 
inequivalent extensions of $A_{4,1}$. There are precisely nine of them. The 
corresponding extensions of $A_{4,2}$ are obtained by the substitution 
(\ref{eq:4.3}). The action of $\hat Y$ on $\{\hat{X}_1,\ldots,\hat{X}_4\}$ is 
represented by the matrix $A$.

\[
\ba{ll}
SA_{5,1} & \hat{X}_1=\paxn+\payn,\ \ \ \ 
\hat{X}_2=t(\paxn+\payn), \\*[2ex]
& \hat{X}_3=\sn \me^{t}(\paxn+\payn),\ \ \ \ 
\hat{X}_4=\sn \me^{-t}(\paxn+\payn),
\\*[2ex]
& \hat{Y}=\pat+a(\xn \paxn+\yn \payn), \\*[2ex]
& \Fn=\xin \fn(\omn),\ \ \ \ \Gn=\frac{\displaystyle\sn}
{\displaystyle\sn-\sd}\etad, \\*[2ex]
& \Kn=\xin \kn(\omn),\ \ \ \ \Pn=\frac{\displaystyle\sn}
{\displaystyle\su-\sn}\etan, \\*[2ex]
& \omn=\xin \me^{-at},\ \ \ \ \su\ne \sn,\ \ \ \ A=\left(
\ba{cccc}
a-1 & 0 & 0 & 0 \\
0 & a+1 & 0 & 0 \\
0 & 0 & a & 0 \\
0 & 0 & -1 & a
\ea
\right).\\
\\
SA_{5,2} & \hat{X}_1=\paxn+\payn,\ \ \ \ 
\hat{X}_2=t(\paxn+\payn), \\*[2ex]
& \hat{X}_3=\sn \cos(t) (\paxn+\payn),\ \ \ \ 
\hat{X}_4=\sn \sin(t)(\paxn+\payn),
\\*[2ex]
& \hat{Y}=\pat+a(\xn \paxn+\yn \payn), \\*[2ex]
& \Fn=\xin \fn(\omn),\ \ \ \ \Gn=\frac{\displaystyle-\sn}
{\displaystyle\sn-\sd}\etad, \\*[2ex]
& \Kn=\xin \kn(\omn),\ \ \ \ \Pn=\frac{\displaystyle-\sn}
{\displaystyle\su-\sn}\etan, \\*[2ex]
& \omn=\xin \me^{-at},\ \ \ \ \su\ne \sn,\ \ \ \ A=\left(
\ba{cccc}
a & 1 & 0 & 0 \\
-1 & a & 0 & 0 \\
0 & 0 & a & 0 \\
0 & 0 & -1 & a
\ea
\right). 
\ea
\]
\[
\ba{ll}
SA_{5,3} & \hat{X}_1=\paxn+\payn,\ \ \ \ 
\hat{X}_2=t(\paxn+\payn), \\*[2ex]
& \hat{X}_3=(\sn + t^{2})(\paxn+\payn),\ \ \ \ 
\hat{X}_4=(\sn t+\frac{t^{3}}{3})(\paxn+\payn),
\\*[2ex]
& \hat{Y}=\pat+a(\xn \paxn+\yn \payn), \\*[2ex]
& \Fn=\xin \fn(\omn),\ \ \ \ \Gn=\frac{\displaystyle2\etad}
{\displaystyle\sn-\sd}, \\*[2ex]
& \Kn=\xin \kn(\omn),\ \ \ \ \Pn=\frac{\displaystyle2\etan}
{\displaystyle\su-\sn}, \\*[2ex]
& \omn=\xin \me^{-at},\ \ \ \ \su\ne \sn,\ \ \ \ A=\left(
\ba{cccc}
a & 0 & 0 & -2 \\
-1 & a & 0 & 0 \\
0 & 0 & a & 0 \\
0 & 0 & -1 & a
\ea
\right).  \\
\\
SA_{5,4} & \hat{X}_1=\paxn+\payn,\ \ \ \ 
\hat{X}_2=t(\paxn+\payn), \\*[2ex]
& \hat{X}_3=\sn t^{\al}(\paxn+\payn),\ \ \ \ 
\hat{X}_4=\sn t^{1-\al}(\paxn+\payn),
\\*[2ex]
& \hat{Y}=t\pat+(a+\frac{1}{2})(\xn \paxn+\yn \payn), \\*[2ex]
& \Fn=t^{-2} \xin \fn(\omn),\ \ \ \ 
\Gn=\al(\al-1)t^{-2}\frac{\displaystyle\sn}{\displaystyle\sn-\sd}\etad,
 \\*[2ex]
& \Kn=t^{-2} \xin \kn(\omn),\ \ \ \ 
\Pn=\al(\al-1)t^{-2}\frac{\displaystyle\sn}{\displaystyle\su-\sn}\etan,
 \\*[2ex]
& \omn=\xin t^{-(a+\frac{1}{2})},\ \ \ \ \su\ne \sn,\ \ \ \ \al\ne 0,1\ , \\  
& A=\diag(a-\al+\frac{1}{2}\, ,\, a+\al-\frac{1}{2}\, ,\, a+\frac{1}{2}\, ,\,  
a-\frac{1}{2}). \\
\\
SA_{5,5} & \hat{X}_1=\paxn+\payn,\ \ \ \ 
\hat{X}_2=t(\paxn+\payn), \\*[2ex]
& \hat{X}_3=\sn t^{1/2} \ln(t)(\paxn+\payn),\ \ \ \ 
\hat{X}_4=\sn t^{1/2}(\paxn+\payn), \\*[2ex] 
& \hat{Y}=t\pat+(a+\frac{1}{2})(\xn \paxn+\yn \payn), \\*[2ex]
& \Fn=t^{-2} \xin \fn(\omn),\ \ \ \ 
\Gn=-\frac{1}{4}t^{-2}\frac{\displaystyle\sn}{\displaystyle\sn-\sd}\etad, 
\\*[2ex]
& \Kn=t^{-2} \xin \kn(\omn),\ \ \ \ 
\Pn=-\frac{1}{4}t^{-2}\frac{\displaystyle\sn}{\displaystyle\su-\sn}\etan, 
\\*[2ex]
& \omn=\xin t^{-(a+\frac{1}{2})},\ \ \ \ \su\ne\sn,\ \ \ \ A=\left(
\ba{cccc}
a & -1 & 0 & 0 \\
0 & a & 0 & 0 \\
0 & 0 & a+\frac{1}{2} & 0 \\
0 & 0 & 0 & a-\frac{1}{2}
\ea
\right). 
\ea
\]
\[
\ba{ll}
SA_{5,6} & \hat{X}_1=\paxn+\payn,\ \ \ \ 
\hat{X}_2=t(\paxn+\payn), \\*[2ex]
& \hat{X}_3=\sn t^{1/2} \cos[\ln(t)](\paxn+\payn),\ \ \ \ 
\hat{X}_4=\sn t^{1/2} \sin[\ln(t)](\paxn+\payn), \\*[2ex] 
& \hat{Y}=t\pat+(a+\frac{1}{2})(\xn \paxn+\yn \payn), \\*[2ex]
&\Fn=t^{-2} \xin \fn(\omn),\ \ \ \ 
\Gn=-\frac{5}{4}t^{-2}\frac{\displaystyle\sn}{\displaystyle\sn-\sd}\etad, 
\\*[2ex]
&\Kn=t^{-2} \xin \kn(\omn),\ \ \ \ 
\Pn=-\frac{5}{4}t^{-2}\frac{\displaystyle\sn}{\displaystyle\su-\sn}\etan, 
\\*[2ex]
& \omn=\xin t^{-(a+\frac{1}{2})},\ \ \ \ \su\ne\sn,\ \ \ \ A=\left(
\ba{cccc}
a & 1 & 0 & 0 \\
-1 & a & 0 & 0 \\
0 & 0 & a+\frac{1}{2} & 0 \\
0 & 0 & 0 & a-\frac{1}{2}
\ea
\right). \\
\\
SA_{5,7} & \hat{X}_1=\paxn+\payn,\ \ \ \ \hat{X}_2=t(\paxn+\payn), \\*[2ex]
& \hat{X}_3=[\sn -\ln(t)](\paxn+\payn),\ \ \ \ 
\hat{X}_4=t[\sn +\ln(t)](\paxn+\payn), \\*[2ex]
& \hat{Y}=t\pat+(a+\frac{1}{2})(\xn \paxn+\yn \payn), \\*[2ex]
& \Fn=t^{-2} \xin \fn(\omn),\ \ \ \ \Gn=-t^{-2} 
\frac{\displaystyle\etad}{\displaystyle\sn-\sd}, \\*[2ex]
& \Kn=t^{-2} \xin \kn(\omn),\ \ \ \ \Pn=-t^{-2} 
\frac{\displaystyle\etan}{\displaystyle\su-\sn}, \\*[2ex]
& \omn=\xin t^{-(a+\frac{1}{2})},\ \ \ \ \su\ne\sn,\ \ \ \ A=\left(
\ba{cccc}
a+\frac{1}{2} & 0 & 1 & 0 \\
0 & a-\frac{1}{2} & 0 & -1 \\
0 & 0 & a+\frac{1}{2} & 0 \\
0 & 0 & 0 & a-\frac{1}{2}
\ea
\right). 
\ea
\]
\[
\ba{ll}
SA_{5,8} & \hat{X}_1=\paxn+\payn,\ \ \ \ \hat{X}_2=t(\paxn+\payn), \\*[2ex]
& \hat{X}_3=\la_{1n}(t)(\paxn+\payn),\ \ \ \ 
\la_{1n}=\sn(t^2+1)^{1/2} \exp[\al \arctan(t)], \\*[2ex]
& \hat{X}_4=\la_{2n}(t)(\paxn+\payn),\ \ \ \  
\la_{2n}=\sn(t^2+1)^{1/2} \exp[-\al \arctan(t)], \\*[2ex]
& \hat{Y}=(t^2+1)\pat+(a+t)(\xn \paxn+\yn \payn), \\*[2ex]
& \Fn=(t^2+1)^{-2} \xin \fn(\omn),\ \ \ \ 
\Gn=(\al^{2}+1)(t^2+1)^{-2}\frac{\displaystyle\sn}
{\displaystyle\sn-\sd}\etad, \\*[2ex]
& \Kn=(t^2+1)^{-2} \xin \kn(\omn),\ \ \ \ 
\Pn=(\al^{2}+1)(t^2+1)^{-2}\frac{\displaystyle\sn}
{\displaystyle\su-\sn}\etan, \\*[2ex]
& \omn=\xin (t^2+1)^{-1/2} \exp[-a \arctan(t)],\ \ \ \ \su\ne\sn,\ \ \ \ 
\al\ne 0, \\*[2ex]
& A=\left(
\ba{cccc}
a-\al & 0 & 0 & 0 \\
0 & a+\al & 0 & 0 \\
0 & 0 & a & 1 \\
0 & 0 & -1 & a
\ea
\right). \\
\\
SA_{5,9} & \hat{X}_1=\paxn+\payn,\ \ \ \ \hat{X}_2=t(\paxn+\payn), \\*[2ex]
& \hat{X}_3=\sn (t^2+1)^{1/2}(\paxn+\payn),\ \ \ \ 
\hat{X}_4=\sn (t^2+1)^{1/2} \arctan(t)(\paxn+\payn), \\*[2ex] 
& \hat{Y}=(t^2+1)\pat+(a+t)(\xn \paxn+\yn \payn), \\*[2ex]
& \Fn=(t^2+1)^{-2} \xin \fn(\omn),\ \ \ \ 
\Gn=(t^2+1)^{-2}\frac{\displaystyle\sn}
{\displaystyle\sn-\sd}\etad, \\*[2ex]
& \Kn=(t^2+1)^{-2} \xin \kn(\omn),\ \ \ \ 
\Pn=(t^2+1)^{-2}\frac{\displaystyle\sn}
{\displaystyle\su-\sn}\etan, \\*[2ex]
& \omn=\xin (t^2+1)^{-1/2} \exp[-a \arctan(t)],\ \ \ \ \su\ne\sn,\ \ \ \ 
\al\ne 0, \\*[2ex]
& A=\left(
\ba{cccc}
a & 0 & 0 & 0 \\
-1 & a & 0 & 0 \\
0 & 0 & a & 1 \\
0 & 0 & -1 & a
\ea
\right).
\ea
\]

In all cases the interaction terms $\Gn$ and $\Pn$ are specified, whereas
 $\Fn$ and $\Kn$ each involve an arbitrary function of one variable 
$\omega_n$. The time dependence of the variable $\omega_n$ and the functions 
$\Fn$ and $\Kn$ depends on the form of the
 generator $\hat{Y}$.

After the substitution (\ref{eq:4.3}) we have altogether 18 five-dimensional 
Lie algebras. No further symmetry generators can be added, at least in the 
nonlinear case 
studied so far.

\section{Nonsolvable symmetry algebras}

Any finite dimensional Lie algebra $L$ that is not solvable is either 
semisimple, or has a nontrivial and unique Levi decomposition

\be
L=S\rhd R,
\label{eq:7.1}
\ee
where $S$ is semisimple and $R$ is the radical, i.e. the maximal solvable 
ideal. The only semisimple Lie algebra that can be realized in terms of the 
vector fields (\ref{eq:2.2}) is actually simple, namely 
$sl(2,\mathbb{R})$.  Up to allowed transformations the realization is unique 
(and given below by 
the operators $\hat{Y}_1$, $\hat{Y}_2$ and $\hat{Y}_3$). 
The determining equations (\ref{eq:2.4}) and (\ref{eq:2.5}) can be used to
 obtain the interaction invariant under the corresponding group 
$SL(2,\mathbb{R})$. Equations (\ref{eq:1.1}) will then be 
invariant under a five-dimensional group that contains the subalgebra 
$A_{2,1}$. We have:

\[
\ba{ll}
NS_{5,1} & \hat{Y}_1=\pat,\ \ \ \ 
\hat{Y}_2=t\pat+\frac{1}{2}(\xn \paxn + \yn \payn),\ \ \ \ 
\hat{Y}_3=t^2\pat+t(\xn \paxn + \yn \payn), 
\\*[2ex]
& \hat{X}_1=\paxn+\payn,\ \ \ \ \hat{X}_2=t(\paxn+\payn), 
\\*[2ex]
& \Fn=\xin^{-3}\, \fn,\ \ \ \ \Gn=\etad^{-3}\, \gn, 
\\*[2ex]
& \Kn=\xin^{-3}\, \kn,\ \ \ \ \Pn=\etan^{-3}\, \pn.
\ea
\]
The Lie algebra $NS_{5,1}$ is isomorphic to the special affine algebra 
$sa\!f\!f(2,\mathbb{R})$. This is the only maximal nonsolvable symmetry 
algebra that occurs.

This completes our analysis of possible symmetries of the system 
(\ref{eq:1.1}) with nonlinear interactions.

\section{Symmetries of linear interactions}

In Sections~3--7 we have excluded the case of linear 
interactions. Let us turn to this case now. We specify equations 
(\ref{eq:1.1}) to be

\be
\ba{rcl}
\ddot{x}_n&=&A_n(t)\, \xin+B_n(t)\, \etad + U_n(t), 
\label{eq:1l} \\*[2ex]
\ddot{y}_n&=&C_n(t)\, \xin+D_n(t)\, \etan + V_n(t)\, .
\label{eq:8.1}
\ea
\ee
The system is still strongly coupled, i.e. the  functions 
$A_n, B_n, C_n, D_n$ are all nonzero. The determining equations reduce to 

\begin{eqnarray}
\ddot{\la}_n-(\mun-\lan)A_n-(\lan-\mud)B_n+(a-\frac{3}{2} 
\dot{\tau})U_n-\tau \dot{U}_n=0,
\label{eq:8.2} \\
\ddot{\mu}_n-(\mun-\lan)C_n-(\lau-\mun)D_n+(a-\frac{3}{2} 
\dot{\tau})V_n-\tau \dot{V}_n=0,
\label{eq:8.3}
\end{eqnarray}
\bea 
2\dot{\tau}A_n+\tau \dot{A}_n=0,
\label{eq:8.4} \\*[2ex]
2\dot{\tau}B_n+\tau \dot{B}_n=0,
\label{eq:8.5} \\*[2ex]
2\dot{\tau}C_n+\tau \dot{C}_n=0,
\label{eq:8.6} \\*[2ex]
2\dot{\tau}D_n+\tau \dot{D}_n=0, 
\label{eq:8.7}\\*[2ex]
\tau=\tau_0+\tau_1+\tau_2\, t^2,
\label{eq:8.7b} 
\end{eqnarray}
since the coefficients of $\xi_n$, $\eta_n$, $\eta_{n-1}$ and $1$ vanish 
separately.

For $A_n(t),\ldots,D_n(t)$ generic, we obtain $\tau=0$ and then only 
equations (\ref{eq:8.2}) and (\ref{eq:8.3}) (with $\tau=0$) survive. These 
equations can be solved in the generic case and we obtain two types of 
symmetries, both just a consequence of linearity.

\begin{enumerate}

\item We take $a=0$ and denote $(\la_{h,n}, \mu_{h,n})$ the general solution
 of the homogeneous equations, i.e. eq.(\ref{eq:8.1}) with $U_n=V_n=0$. The 
vector field is

\be
\hat{X}_h=\la_{h,n}(t)\, \paxn + \mu_{h,n}(t)\, \payn.
\label{eq:8.8}
\ee

\item For $a\ne 0$ we choose $a=-1$ and denote some chosen particular 
solution of the inhomogeneous system (\ref{eq:8.1}) $(\la_{p,n}, \mu_{p,n})$. 
The vector field is 

\be
\hat{X}_p=[\xn-\la_{p,n}(t)]\paxn + [\yn-\mu_{p,n}(t)]\payn\, .
\label{eq:8.9}
\ee
In particular, if we have $U_n=V_n=0$, then we take 
$\la_{p,n}=\mu_{p,n}=0$ in eq.(\ref{eq:8.9}).

\end{enumerate}

The symmetry (\ref{eq:8.8}) only means that we can add any solution of the
 homogeneous equations to a solution of eq.(\ref{eq:8.1}). The symmetry 
(\ref{eq:8.9}) corresponds to the fact that a solution of the homogeneous 
system can be multiplied by a constant.

Let us now assume that a further symmetry generator exists. It is of the form 
(\ref{eq:2.2}) with $\tau(t)$ as in eq.(\ref{eq:8.7b}). Allowed transformations
 can be used to transform $\tau$ into one of four cases. Let us consider them 
separately.

\subsection*{a) $\tau=0$}
No symmetries beyond the generic ones are obtained.

\subsection*{b) $\tau=1$}
Using allowed transformations we simplify the additional 
vector field into

\be
\hat{T}=\pat+a(\xn\, \paxn+\yn\, \payn).
\label{eq:8.10}
\ee
The determining equations restrict the time dependence of the coefficients in
 eq.(\ref{eq:8.1}) and the system reduces to

\be
\ba{rcl}
\ddot{x}_n&=&f_n\, \xin+g_n\, \etad + u_n \me^{at}, 
\\*[2ex]
\ddot{y}_n&=&k_n\, \xin+p_n\, \etan + v_n \me^{at}.
\label{eq:8.11}
\ea
\ee

\subsection*{c) $\tau=t$}
The additional vector field and invariant equations are reduced
 to

\be
\hat{D}=t\, \pat+(a+\frac{1}{2})(\xn\, \paxn+\yn\, \payn),
\label{eq:8.12}
\ee
\be
\ba{rcl}
\ddot{x}_n&=&\frac{\displaystyle f_n}{\displaystyle t^2} \xin+
\frac{\displaystyle g_n}{\displaystyle t^2}\, \etad 
+ u_n t^{a-\frac{3}{2}}, 
\\*[2ex]
\ddot{y}_n&=&\frac{\displaystyle k_n}{\displaystyle t^2}\, \xin+
\frac{\displaystyle p_n}{\displaystyle t^2}\, \etan 
+ v_n  t^{a-\frac{3}{2}}\, .
\label{eq:8.13}
\ea
\ee

\subsection*{d) $\tau=t^2+1$}

The additional vector field and invariant equations are

\be
\hat{C}=(t^2+1)\pat+(a+t)(\xn \paxn+\yn \payn),
\label{eq:8.14}
\ee
\be
\ba{rcl}
\\
\ddot{x}_n&=&\frac{\displaystyle f_n}{\displaystyle (t^2+1)^{2}} \xin+
\frac{\displaystyle g_n}{\displaystyle (t^2+1)^{2}}\, \etad 
+\frac{\displaystyle u_n}{\displaystyle (t^2+1)^{3/2}} \exp[a\arctan(t)],
\label{eq:tc1taunn}\\*[2ex]
\ddot{y}_n&=&\frac{\displaystyle k_n}{\displaystyle (t^2+1)^{2}}\, \xin+
\frac{\displaystyle p_n}{\displaystyle (t^2+1)^{2}}\, \etan 
+ \frac{\displaystyle v_n}{\displaystyle (t^2+1)^{3/2}} \exp[a\arctan(t)]\, .
\label{eq:8.15}
\ea
\ee

In all cases $\fn$, $\gn$, $\kn$, $\pn$, $u_n$ and $v_n$ are independent of
 $t$. No further symmetries exist for any of the interactions (\ref{eq:8.11}), 
(\ref{eq:8.13}) or (\ref{eq:8.15}).

\section{Conclusions}

Let us sum up the results obtained above.

For nonlinear interactions the symmetry algebra is at most five-dimensional. 
The following cases occur.

\begin{enumerate}

\item The nonsolvable algebra $NS_{5,1}$ of Section~7. The dependence of the 
right hand 
side of eq.(\ref{eq:1.1}) on $\xin$ and $\etan$ is 
completely specified by an inverse cube relation. The dependence on the 
discrete variable $n$ remains arbitrary. The interactions are time 
independent.

\item The solvable Lie algebras with Abelian nilradicals 
$SA_{5,1},\ldots,SA_{5,9}$ (and $SA_{5,10},\ldots,SA_{5,18}$ by the 
substitution (\ref{eq:4.3})) of Section~6.2. The interactions are all 
``semilinear''. By 
this we mean that the dependence on one variable $\etan$ is specified to be 
linear, whereas the dependence on $\xin$ remains arbitrary (and vice versa 
for $SA_{5,10},\ldots,SA_{5,18}$). The time dependence of the nonlinear terms 
in the interaction depends crucially on the form of the nonnilpotent element 
$\hat{Y}$.

Any attempt to enlarge the symmetry algebra by further elements leads to 
linear interactions.

\item The nilpotent five-dimensional Lie algebras $N_{5,1}$ and the related 
algebra $N_{5,2}$ of Section~4. For $N_{5,1}$ the interaction is again 
semilinear with 
$\Gn$ and $\Pn$ linear in $\etad$ and $\etan$, respectively, and $\Fn$ and 
$\Kn$ arbitrary functions of $\xin$ (and vice versa for $N_{5,2}$). The 
interaction is time independent.

\item Four-dimensional maximal symmetry algebras are either Abelian, or 
solvable with the Heisenberg algebra as a nilradical. For $A_{4,1}$ and 
$A_{4,2}$ the interaction is semilinear with an 
arbitrary time dependence in the nonlinear terms. For $SN_{4,1}$, $SN_{4,2}$ 
and $SN_{4,3}$ the dependence on $\xin$ and $\etan$ is completely specified as 
being monomial, logarithmic or exponential, respectively. There is no 
time dependence.

\item A three-dimensional maximal symmetry algebra is either nilpotent, or 
solvable with an Abelian nilradical. For $N_{3,1}$, the Heisenberg 
algebra, the interaction is time independent, otherwise arbitrary. The model,
 studied by Campa {\em et al} \cite{1}, namely

\[
\ba{rclcrcl}
\Fn(\xin)&=&\frac{1}{M_1}(k_1\xin+\ep \b_1 \xin^2) &,& 
\Kn(\xin)&=&-\frac{M_1}{M_2}F_n(\xin),
\\*[2ex]
\Gn(\etad)&=&-\frac{1}{M_1}(k_2\etad+\ep \b_2 \etad^2) &,& 
\Pn(\etan)&=&-\frac{M_1}{M_2}G_{n+1}(\eta_{n+1}),
\ea
\]
is of this type. For $SA_{3,1}$, $SA_{3,2}$ and $SA_{3,3}$ the interactions 
involve four arbitrary functions of one variable. The interaction is entirely 
specified by the element $\hat{Y}$.

\item As mentioned above, the general interaction in eq.(\ref{eq:1.1}) is 
invariant under the group of global translations and Galilei transformations, 
corresponding to the algebra $A_{2,1}$ of eq.(\ref{eq:2.17}).

\end{enumerate}

The symmetries found in this article can be used to perform symmetry reduction
 on one hand, and to obtain new solutions from known ones, on the other. 

Let us look at the example of algebra $NS_{5,1}$. The system (\ref{eq:1.1}) in
 this case reduces to

\be
\ddot{x}_n=\frac{\fn}{\xin^3}+\frac{\gn}{\etad^3},\ \ \ \ \ \ 
\ddot{y}_n=\frac{\kn}{\xin^3}+\frac{\pn}{\etan^3}.
\label{eq:9.2}
\ee
The algebra $sl(2,\mathbb{R})$ has three inequivalent one-dimensional 
subalgebras, namely $\hat{Y}_1$, $\hat{Y}_2$ and $\hat{Y}_3+\hat{Y}_1$. Each 
of them can be used to reduce the system (\ref{eq:9.2}) to a system of two 
difference equations. Let us look at the three individual cases separately.

\subsection*{Subalgebra $\hat{Y}_1$}

This algebra leads to stationary solutions. We have

\be
x_n=x_{n,0},\ \ \ \ \ \ y_{n}=y_{n,0}
\label{eq:9.3}
\ee
and hence
\be
\xi_{n,0}=\left(-\frac{f_n}{g_n}\right)^{1/3}\eta_{n-1,0}=
\left(-\frac{k_n}{p_n}\right)^{1/3}\eta_{n,0}.
\label{eq:9.4}
\ee

\subsection*{Subalgebra $\hat{Y}_2$}

The reduction formulas in this case are

\be
\xn=x_{n,0}\sqrt{t},\ \ \ \ \ \ \yn=y_{n,0}\sqrt{t}
\label{eq:9.5}
\ee
and the recursion relations are

\be
-\frac{x_{n,0}}{4}=\frac{\fn}{\xi_{n,0}^3}+\frac{\gn}{\eta_{n-1,0}^3},
\ \ \ \ \ \ -\frac{y_{n,0}}{4}=\frac{\kn}{\xi_{n,0}^3}+\frac{\pn}
{\eta_{n,0}^3}.
\label{eq:9.6}
\ee

\subsection*{Subalgebra $\hat{Y}_3+\hat{Y}_1$}

We put

\be
\xn=x_{n,0}\sqrt{t^2+1},\ \ \ \ \ \ \yn=y_{n,0}\sqrt{t^2+1}
\label{eq:9.7}
\ee
and obtain the recursion relations

\be
x_{n,0}=\frac{\fn}{\xi_{n,0}^3}+\frac{\gn}{\eta_{n-1,0}^3},\ \ \ \ \ \ 
y_{n,0}=\frac{\kn}{\xi_{n,0}^3}+\frac{\pn}{\eta_{n,0}^3}.
\label{eq:9.8}
\ee

In all three cases we can express $\xin$ in terms of $\etan$ and obtain a two 
term recursion relation for $\etan$. These can be solved, but we will not go 
into the details here.

\section*{Acknowledgments}

We thank professor D.~Levi for interesting discussions. The research of S.L. 
was successively supported by a FCAR Doctoral Scholarship and by an NSERC 
Postdoctoral Fellowship. The research of P.W. was partly supported by NSERC of
 Canada and FCAR du Qu\'ebec. The work reported here was
 started while all three authors were visiting the CIF in Cuernavaca, Mexico.
 We thank T.~ Seligman for his hospitality there. The final version was 
written while P.W. was visiting the Faculty of Nuclear Sciences and Physical 
Engineering of the Czech Technical University in Prague. He thanks Professors 
M.~Havlicek and J.~Tolar for their hospitality.

\newpage

$$
\xymatrix{
\ar@{.}'[r]&
\ar@{-}'[rrrr]\ar@{-}'[rr]^*+{y_{n-1}}&{\bullet}\ar@{~}[rr]
\ar@{-}'[rrrr]^*+{x_{n}}&&{\circ}\ar@{~}[rr]
\ar@{-}'[rrrr]&\ar@{-}'[rr]^*+{y_{n}}&{\bullet}
\ar@{~}[rr]
\ar@{-}'[rrr]&\ar@{-}'[rr]^*+{x_{n+1}}&{\circ}
&\ar@{.}'[r]&
}
$$

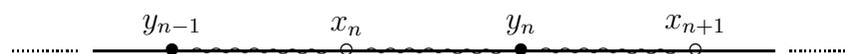
\noindent{\bf \mbox{Figure $1$.}} Interactions between atoms of type $X$ and
 $Y$ along a molecular chain.

\end{document}